\documentclass[prb,twocolumn,showpacs]{revtex4}
\usepackage{graphicx}%
\usepackage{dcolumn}
\usepackage{amsmath}
\begin{document}

\title{Shot-noise in Transport and Beam Experiments}
\author{U. Gavish, Y. Levinson and Y. Imry}
 \affiliation{%
Department of Condensed Matter Physics, Weizmann Institute of Science, Rehovot, 76100, Israel}%

\date{November 2001}

\begin{abstract}
Consider two Fermi gases with the same {\it average} currents: a
{\it transport gas}, as in solid-state experiments where the
chemical potentials of terminal 1 is $\mu+eV$ and of terminal 2 and 3 is $\mu$,
 and a {\it beam}, i.e., electrons entering only from
terminal 1 having energies between $\mu$ and $\mu+eV$. By
expressing the current noise as a sum over single-particle
transitions we show that the temporal current {\it fluctuations} are very
different: The beam is noisier due to allowed single-particle
transitions into empty states below $\mu$. Surprisingly, the
correlations between terminals 2 and 3 are the same.
\end{abstract}


\maketitle

The subject of quantum shot noise \cite{KhlusLesovikButtiker,HeiblumGlattli,Prober}
has recently been of major interest, for example, due to the possibility
 to observe different quasi-particle charges of
the carriers \cite{GlatMotFraction}. Attempts to examine analogies with
Hanbury-Brown and Twiss \cite{HBT electrons,HBT electrons2,tonomura} correlations deserve
particular attention. In 1918 Schottky \cite{schottky} observed that one
contribution (called shot-noise) to the noise in currents flowing in vacuum tubes was due to the
discreteness of the electrons. Presently, most experiments on electronic
noise (an exception is, e.g., Ref. \onlinecite{tonomura}) are performed in a
degenerate Fermi gas and not in vacuum beams. Despite that, they are
often analyzed in a similar fashion to vacuum beams \cite{HBT electrons,HBT electrons2,liu}.
Below it is shown that this point of view is not justified
 since the temporal noise correlators in a given terminal
are substantially different in beams and degenerate Fermi systems
(surprisingly the correlations between different terminals turn out to be same).
  To show this, we shall apply our approach \cite{Gavish Levinson Imry} of viewing
noise as the radiation (or excitations of a detector
\cite{LesovikLoosen}) produced by the current fluctuations.

We consider current fluctuations for two types of Fermi gases in a
ballistic conductor which consists of three single-channel arms connected
to an elastic scatterer (fig.1), assuming zero temperature and
non-interacting electrons. The scattering state with energy
$\epsilon_n=k^2/2m,$ corresponding to a wave that is incoming on arm
$\alpha$, partially reflected back into it and partially transmitted into
the other arms, is: $\varphi_n(x_\beta)= L^{-1/2}
[\delta_{\alpha\beta}e^{-ikx_{\beta}}+s_{\beta\alpha}(k)e^{ikx_\beta}]$.
Here $n\equiv (\alpha,k)$ with $k>0$, $ s_{\alpha \beta}$ is the
scattering matrix, $\alpha , \beta =1,2,3$, $L$ is a normalization length,
$m$ the electron mass and $x_{\beta}$ the distance of a point on arm
$\beta$ from the scatterer. To specify that a state $\varphi_{n}$
comes from terminal $\alpha,$ we shall write $n \in \alpha.$

We compare the current fluctuations in two many-body states
(fig.2), the {\it transport gas}:
\begin{eqnarray}
|transport \rangle \equiv &\prod&  \hat a_n^{\dag} |vacuum \rangle,
\nonumber\\ &_{n \in 1 ;\ \mu \leq \epsilon_n \leq \mu+eV}& \nonumber \\ &_{n \in
1,2,3 ;\ \epsilon_n \leq \mu}&
\label{deftransportgas}
\end{eqnarray}
\vspace{-0.2cm}
and the {\it beam}:
\begin{eqnarray}
 |beam \rangle \equiv &\prod& \hat a_n^{\dag}|vacuum \rangle.
  \nonumber \\
 &_{n \in 1 ;\ \mu \leq \epsilon_n \leq \mu+eV}&
\label{defbeam}
\end{eqnarray}
 where $\hat{a}_n$ and  $\hat{a}_n^{\dag}$
 are the annihilation and creation operators of the $\varphi$'s. In the  transport gas all the
$\varphi_n$'s are occupied up to an energy $ \mu$ if $n\in 2,3$ and up to
$\mu +eV$ if  $n\in 1$. In the beam the only occupied states are all those
coming in on arm 1, which are in the energy range  $[\mu,\mu+eV].$
 It is  assumed  that $eV\ll\mu.$

 The current operator on the arm  $\beta$ is $\hat{j}(x_\beta)=
 -(ie/2m) \sum_{nn'} \hat{a}_{n}^{\dag} \hat{a}_{n'} \varphi_{n}
 ^\ast \nabla_{\beta}\varphi_{n'}+h.c.$.
We assume that the {\it measured} current is the average
\begin{equation}
\hat{J_\beta}\equiv \frac{1}{L_0}\int_{L_0}dx_{\beta}\hat{j} (x_\beta)
\label{J}
\end{equation}
over a segment  $L_0$  far away from the scatterer (fig.1) which
satisfies: $L_0 k_F\gg 1$ and $\omega L_0m/k_F \ll 1,$ where $k_F \equiv
\sqrt{2m\mu}$, and $\omega$ is the frequency of the measured noise which
is assumed to satisfy $\omega\ll\mu$. These conditions ensure that the current
correlations are independent of the length and position of
the segment $L_0,$ and thus will have no spatial dependence, which is not
addressed in experiments.

We consider correlators of  the current fluctuations in the frequency domain:
\begin{equation} S_{\alpha
\beta}(\omega)\equiv \int_{-\infty}^{\infty} e^{i\omega t}\langle i|\hat
J_{\alpha}(0) \hat J_{\beta}(t)|i\rangle dt , \label{defS}
\end{equation} where $\hat J_{\alpha}(t)=e^{iHt}\hat J_{\alpha}e^{-iHt}
$ is the Heisenberg representation of $\hat{J}_\alpha$ and $|i \rangle =|beam
\rangle ,|transport \rangle$. There is an alternative definition as a
Fourier transform of the symmetrized correlator $(1/2)\langle i|\hat J_{\alpha}
(0)\hat J_{\beta}(t) + \hat J_{\beta}(t) \hat J_{\alpha}(0) |i\rangle $.
We use the non-symmetrized version since, following ref.
\onlinecite{LesovikLoosen}, we showed \cite{Gavish Levinson Imry} that at
least for $\alpha=\beta$ and for some types of noise detection, it is Eq.(\ref{defS})
which gives
the measured noise if the detector is cold enough.
\begin{figure}
     \begin{center}
         \includegraphics[height=0.9in,width=1.90in,angle=0]{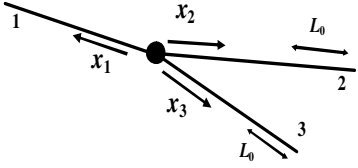}
\vspace{-0.1cm}
         \caption{Three leads connected to an elastic scatterer}
     \end{center}
 \end{figure}

\begin{figure}
     \begin{center}
         \includegraphics[height=3.1in,width=3.2in,angle=0]{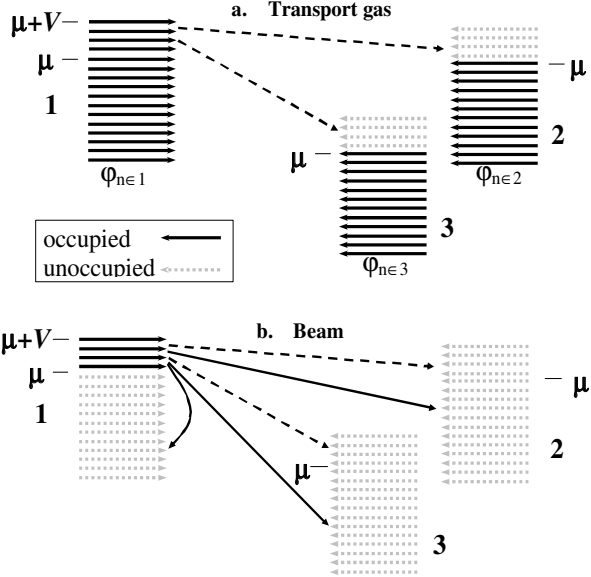}
         \vspace{-0.5cm}
         \caption{The occupations and possible transitions for $\omega >0$. Short horizontal arrows represent scattering states.  }
         \vspace{-0.7cm}
     \end{center}
\end{figure}
Following the ideas in neutron-scattering theory introduced by
Van-Hove\cite{vanHove}  we insert a complete set of eigenstates into
Eq.(\ref{defS}) and get after a short manipulation:
\begin{equation}
S_{\alpha \beta}(\omega)=2\pi\sum_f \langle i| \hat J_{\alpha}(0) |f
\rangle \langle f| \hat J_{\beta}(0) |i \rangle \delta (E_{i} - E_{f} -
\omega), \label{Svanhovemanybody}
\end{equation}
where  $E_i$ and $ E_f$ are the energies of $|i \rangle$ and $|f \rangle$.
The non-diagonal element  $\langle i| \hat J_\alpha(0) |f \rangle $ is
nonzero only if $|f \rangle$ differs from $|i \rangle$ by moving one
particle from an occupied state, $\varphi_n,$ to a  previously unoccupied
state, $\varphi_{n'},$ i.e., $|f \rangle$ is of the form $ \hat{a}_{n'}^{\dag}
\hat{a}_n |i \rangle$ (up to a fermionic factor $c=\pm 1$, that will play
no role below.) The term with the diagonal element $\langle i| \hat
J_\alpha(0) |i \rangle $ which is the average current, $I_\alpha(V)$, on arm
$\alpha$, yields a term $\sim \delta (\omega)$. In what follows we
consider only $\omega \not= 0$ and therefore neglect this term. In
experiments the integration in Eq.(\ref{defS}) is limited by the sampling
time of the experiment, $T_s$, and as a result  $\delta (\omega)$ is
smoothed into a peak with a width of $\simeq 1/T_s$ which means that the
condition $\omega \not=0$ actually is $\omega T_s\gg 1$. We therefore
have:
\begin{equation}
S_{\alpha \beta}(\omega)=2\pi \sum_{nn'} J_{\alpha,nn'}J_{\beta,nn'}^\ast
\delta (\epsilon_n-\epsilon_{n'} - \omega), \label{Svanhove}
\end{equation}
where $J_{\alpha,nn'} \equiv \langle i| \hat J_{\alpha}(0)
\hat{a}_{n'}^{\dag} \hat{a}_n |i \rangle$, and where now the summation over
$n$ and $n'$ is over all {\it single-particle} states $\varphi _n$ and $\varphi
_{n'}$ which are occupied and unoccupied, respectively, in $|i\rangle$.
The auto-correlator is
\begin{eqnarray}
 S_{\alpha \alpha}(\omega)=2\pi
\sum_{nn' }|J_{\alpha, nn'}|^2 \delta (\epsilon_{n}-\epsilon_{n'} -
\omega). \label{Svanhove-auto}
\end{eqnarray}
 When the system is coupled to a measuring device (e.g., some circuit
or an electro-magnetic field) through a small term linear in $\hat{J}_{\alpha}$,
$S_{\alpha \alpha}(\omega)$ is a sum over {\it
single-particle transitions}, the probability of each given by the
Fermi golden rule, between an initial $\varphi_{n}$ and a final $\varphi_{n'}$.
 (The cross-correlator,  Eq.(\ref{Svanhove}) for
$\alpha \not=\beta$,
 should not be viewed similarly since then $J_{\alpha,nn'}J_{\beta,nn'}^\ast$ is not a transition
amplitude squared). Via these transitions energy is transferred between
the system and the measuring device: terms with $
\epsilon_{n}>\epsilon_{n'}$ (one particle goes down in energy) describe
transitions in which an energy of  $\omega = \epsilon_{n} - \epsilon_{n'}
>0$ is transferred  from the system to the measuring device, while terms
with $ \epsilon_{n'}>\epsilon_{n}$ (one particle goes up) describe
transitions in which an energy of $-\omega = \epsilon_{n'} - \epsilon_{n}
>0$ is transferred  from the measuring device to  the system.
When $\omega>0$, only the first type of terms will
remain and $S_{\alpha \alpha}(\omega)$ will be the emission spectrum while
$S_{\alpha \alpha}(-\omega)$ is the absorption spectrum. Thus we conclude
that when $\alpha=\beta$ there will be emission of noise at frequency
$\omega$   {\it if and only if\/}  there exist occupied and unoccupied
states in $|i \rangle ,$  $\varphi_{n}$ and $\varphi_{n'}$, with
$|J_{\alpha,nn'}|^2 \not= 0$ and $\omega = \epsilon_n - \epsilon_{n'} >0.$
For $\alpha \not= \beta$, this is not necessarily so, since
the terms in Eq.(\ref{Svanhove}) are complex and may cancel.

Now let us compare the current and its fluctuations in the transport gas
and the beam, considering the arms 2 and 3. The  average  currents in both
systems are defined only by states in the energy window $[\mu, \mu +eV]$
and are the same: for $\beta =2,3$ one finds $I_{\beta}(V)=e^2T_\beta
V/(2\pi)$ both for $ |i\rangle = |beam\rangle$ and $|i\rangle =|transport
\rangle$, where $T_\beta \equiv |s_{\beta 1} |^2$. (For simplicity we
neglected the energy dependence of the transmission).
 By calculating the emission spectrum ($\omega
>0$) we now show that the current
fluctuations may differ.  Rewriting Eqs.(\ref{Svanhove}) and
(\ref{Svanhove-auto}) for $\omega>0,$ taking into account the
energy conservation and the different occupations in the states
Eqs.(\ref{deftransportgas}) and (\ref{defbeam}), one has for the
transport gas:\vspace{-.01cm}
\begin{eqnarray}
S_{\alpha \beta}^{tr}(\omega)
=2\pi\sum J_{\alpha,nn'}J_{\beta,nn'}^\ast\delta ( \epsilon_n-\epsilon_{n'} - \omega)
\nonumber \\ _{n \in 1,n'\in 2,3} \
_{\mu<\epsilon_{n'}<\epsilon_n<\mu+eV}\qquad\qquad\qquad
 \label{Stransport}
\end{eqnarray}
and for the beam:
\begin{equation}
S_{\alpha \beta}^{b}(\omega)= S_{\alpha \beta}^{tr}(\omega)+ \sum
_{n \in 1 ;\   \mu< \epsilon_n< \mu +\min(\omega, eV)}
 S_{\alpha \beta}^{(n)}(\omega).
 \label{Sbeam}
\end{equation}
Here $S_{\alpha \beta}^{(n)}(\omega)$ corresponds to the
correlator in the state ${\hat{a}_n}^{\dag} |vacuum \rangle$, which is a beam
with a {\it single\/} particle, in a state $\varphi _n$:
\begin{equation}
S_{\alpha \beta}^{(n)} (\omega)= 2\pi \sum_{n'\in 1,2,3 }
J_{\alpha,nn'}J_{\beta,nn'} ^\ast  \delta ( \epsilon_{n} -\epsilon_{n'} -
\omega). \label{Sn}
\end{equation}
$S_{\alpha \beta}^{tr}(\omega)$ contains transition amplitudes between
occupied states in the energy window $[\mu, \mu +eV]$ to lower empty
states inside the same energy window. These transitions, shown by dashed
arrows in figs. 2a, and 2b, are possible both in the transport gas and the
beam and therefore $S_{\alpha \beta}^{tr}(\omega)$ appears also in
Eq.(\ref{Sbeam}). Contrary to the first one, the second term in
Eq.(\ref{Sbeam}) contains transition amplitudes between occupied states in
the energy window $[\mu, \mu +eV]$ to empty states below $\mu$ (long solid
arrows in fig. 2b), transitions which are allowed only in the beam.
Writing this term as a sum of single-particle correlators was possible
since in the beam all the levels below $\mu$ are empty
so the sum runs over all possible values of $n'$ with a given energy,
unlike in Eq.(\ref{Stransport}) for the transport gas where $n'\notin 1$.

Now, the current matrix element is given by:
\begin{eqnarray}
\label{melna}
\nonumber
\langle i|\hat{j}_{\alpha}\;\hat{a}^{\dag}_{n'}\hat{a}_{n}|i\rangle=
(c/2)\left[(k'+k)e^{i(k-k')x_{\alpha}}s_{\alpha 1}(k)^{\ast}\times\right.
\\
 \left.
s_{\alpha\gamma'}(k')+(k'-k)s_{\alpha 1}(k)^{\ast}\delta _{\alpha \gamma'}
e^{-i(k+k')x_{\alpha}}\right]
\end{eqnarray}
for $n=(k,1)$ occupied and $n'=(k',\gamma')$ empty. $c=\pm 1$, as above. Performing the average
as defined in Eq.(\ref{J}) and using the conditions for $L_{0}$, we obtain
\begin{eqnarray}
\label{melav} J_{\alpha,nn'}=(c/2) (k+k')s_{\alpha
1}(k)^{\ast}s_{\alpha\gamma'}(k').
\end{eqnarray}
This matrix element has no spatial dependence because the fast oscillating
term in Eq.(\ref{melna}) vanished while the slow oscillating one is
constant within $L_{0}$. Inserting Eq. (\ref{melav}) into
Eq.(\ref{Svanhove}), transforming the sums over $k$ and $k'$ into
integrals, integrating using the condition $\omega,eV \ll \mu$, using the unitarity of the scattering matrix, $\sum_{\gamma}
s_{\alpha \gamma} s_{\beta \gamma}^\ast=\delta_{\alpha \beta}$, one gets \cite{KhlusLesovikButtiker}
for $\alpha , \beta =2,3 $, and $\omega >0$:
\begin{eqnarray}
S_{\alpha \beta}^{tr}(\omega)=\frac{e^2}{2\pi} T_{\alpha}(\delta_{\alpha
\beta}-T_{\beta})(eV-\omega ) \theta (eV- \omega)&,
\label{Stransportgas-final}
\end{eqnarray}
where $\theta$ is the Heaviside step-function.
 Similarly, using Eq.(\ref{melav}) in Eq.(\ref{Sn}) one gets
for the single-particle correlator (see discussion below), for $\alpha , \beta =2,3$:
\begin{equation}
 S_{\alpha \beta}^{(n)}(\omega) = \delta_{\alpha \beta}eI_{n,\alpha},
\label{Sn-final}
\end{equation}
 where $I_{n,\alpha}=eT_\alpha (k/m)(1/L)$ is the average current on arm
$\alpha =2,3$ of a single particle in the state $n=(k,1)$.

Substituting  Eq.(\ref{Sn-final}) in Eq.(\ref{Sbeam}), one gets for
$\alpha , \beta =2,3 $:
\begin{equation}
 S_{\alpha \beta}^{b}(\omega)= S_{\alpha \beta}^{tr}(\omega)
+\delta_{\alpha \beta}eI_{\alpha}(\omega),
\label{Sbeam-final}
\end{equation}
where $I_{\alpha}(\omega)$ is the average current of the electrons in the energy window
$[\mu,\mu +\min(\omega, eV)]$:
\begin{equation}
I_{\alpha}(\omega)\equiv \sum_{n \in 1;
 \ \mu< \epsilon_n < \mu+\min(\omega,eV)}
I_{n, \alpha}= I_\alpha (V) \frac {\min(\omega ,eV)}{eV}.
\label{Icl}
\end{equation}

Eq.(\ref{Sbeam-final}) is our main result and it demonstrates that
although the {\it average} currents in arms 2 and 3 in the beam and the
transport gas are the same, the current {\it fluctuations} in these arms
generally differ. The beam has much more noise: e.g., for
$\alpha=2,3$ the auto-correlation spectra of the transport state
$S^{tr}_{\alpha\alpha}(\omega)$ has an upper cutoff at $\omega=eV$, but the
auto-correlation spectra of the beam $S^{b}_{\alpha\alpha}(\omega)$  has
no such cutoff. The spectra $S^{b}_{\alpha\alpha}(\omega)$ at $\omega>eV$
is given by the extra second term in in Eq.(\ref{Sbeam-final}).
Interestingly, this term is identical
to the result for a beam of uncorrelated (Poissonian) classical particles \cite{{S^n}} which carries an average
current given by Eq.(\ref{Icl}).
Surprisingly, the cross-correlation, $S_{23}(\omega)$ is identical in
the beam and the transport gas, since this term vanishes for $\alpha\not= \beta$.

According to Eq.(\ref{Sbeam-final}) and Eq.(\ref{Icl}) $S^{b}_{\alpha \alpha}(\omega)$ and
$S^{tr}_{\alpha\alpha}(\omega)$ start to differ substantially for $\omega$ of order $eV$. The
measurement in Ref. \onlinecite{HBT electrons2} is consistent with Eq.(\ref{Stransportgas-final})
but since it is performed at $\omega\ll eV$, it can not distinguish
between $S^{tr}_{\alpha\alpha}(\omega)$ and $S^{b}_{\alpha\alpha}(\omega)$.
In Ref. \onlinecite{HBT electrons} (see particularly Fig.3)
it is claimed that the cross-correlation are measured in the time domain.
 The function that was obtained via this measurement has characteristic time-scale of $\sim 100ns$
 which, in the frequency domain, corresponds to $10MHz$.
However, in both Eq.(\ref{Sbeam-final}) and Eq.(\ref{Icl}) the only characteristic
frequency scale is of the order of $eV\approx 10^{5}MHz$
 (estimated for $\sim 30nA$ and transmission of order 1), that is many
 orders of magnitude larger. Thus, the results in Ref. \onlinecite{HBT electrons} are not consistent
 with ours.

The simple case of a two-terminal device is obtained from the
from Eq.(\ref{Sbeam-final}) by taking $s_{13}=0$.
In this case there is only one independent
correlator, since $S_{11}(\omega)=S_{22}(\omega)=-S_{12}(\omega)\equiv
S(\omega)$. All these correlators are different for the transport gas and
the beam. Denoting $I$ as the average current in the device, one gets:
\begin{eqnarray}
S^{b}(\omega)-S^{tr}(\omega)=eI\frac{\min(eV,\omega)}{eV},
\end{eqnarray}

We now explain the classical form of the extra term in Eq.(\ref{Sbeam-final}).
This term contains transition amplitudes from states in the energy window
 $[\mu, \mu +min(\omega,eV)]$ to states below $\mu$
 (see the second term in Eq.(\ref{Sbeam}) and Eq.(\ref {Sn})) .
  Since {\it all} final states are empty the quantum statistics plays no role.
 So, with no interactions and no statistics, the particles in this energy window are independent.
For \textit{independent} particles (classical or quantum), the correlator is a
sum of their single-particle correlators: $S_{\alpha \beta}=\sum_n S_{\alpha\beta}^{(n)}$
(see the second term in Eq.(\ref{Sbeam}) and Ref.\onlinecite{S^n}).
Since the classical single-particle correlator is identical to its quantum counter-part according to Eq.(\ref{Sn-final})
and Ref.\onlinecite{S^n}, the contribution of the particles in the above energy window has a classical form.

It remains to understand why the quantum and classical single-particle correlators
are equal.
This is due to the averaging in Eq.(\ref{J}),
the unitarity of the scattering matrix and the assumption $\omega\ll \mu$.
Here we will explain in detail only the role of the averaging in the vanishing of the cross-correlation:
Before averaging, the single-particle temporal cross-correlator is
\begin{eqnarray}
\nonumber
\langle \varphi_{1,k}|\hat j(x_2,0) \hat j(x_3,t) |\varphi_{1,k} \rangle=
\frac{s_{21}s_{31}}{4L}e^{i\epsilon_{1,k}t} \times\\
 e^{-ik(x_2-x_3)}(-i\partial_{x_2}+k)(i\partial_{x_3}+k)
\langle x_2| e^{-iHt}  |x_3 \rangle
\label{jj(t)}
\end{eqnarray}
where $|\varphi_{1,k} \rangle \equiv {\hat{a}_{1,k}}^{\dag} |vacuum \rangle$, and where

\begin{equation}
\langle x_2| e^{-iHt)}|x_3 \rangle= \frac{s_{23}}{\sqrt{2\pi t}}exp(-i\frac{(x_2+x_3)^2}{2mt}+i\frac{\pi}{4})
\label{xHx}
\end{equation}
is the (generally nonzero) propagator from $(x_2,0)$ to $(x_3,t)$, for $t>0$. For simplicity, we assume the $s_{\alpha,\beta}$'s are real and
$k$-independent. The above correlator,
Eq.(\ref{jj(t)}), is generally different from zero (in contrast to its classical counter part).
However, when applying the spatial averaging each of its terms becomes proportional to a new type of propagators:
 of a \textit{wave-packet} around momentum $k>0$
which is localized in a segment of size $L_0$ around $x_2$ on arm 2 into a similar wave-packet around
 $x_3$ on arm 3.
 This is so because the factor $e^{-ik(x_2-x_3)}\langle x_2| e^{-iHt}|x_3 \rangle$ in
Eq.(\ref{jj(t)}) turns upon averaging into:
\begin{equation}
\left( \int_{L_0} \frac{dx_2}{L_0} e^{-ikx_2} \langle x_2|\right)  e^{-iHt}
 \left(\int_{L_0} \frac{dx_3}{L_0} e^{ikx_3} |x_3 \rangle \right)
\label{JJ(t)}
\end{equation}
where each integral is a wave-packet of the form described above.
All the four terms in Eq.(\ref{jj(t)}), become after averaging,
proportional to propagators of similar though more complicated form.
\textit{These} propagators vanish in the limit $kL_0\gg 1$,
causing the quantum single-particle cross-correlator of the \textit{average} current
to vanish, similarly to the classical one. This vanishing has a physical meaning:
 if a particle is at a point $x_2$ on arm 2 it has, due to Heisenberg-principle,
large momentum uncertainty and thus, although it is already on arm 2, a possibility to
return and be scattered into arm 3 and reach $x_3$. However, when it is spread out in a segment
of size $L_0$ around $x_2$ which is much larger than the inverse of the average momentum, its momentum uncertainty
is not enough to allow it to return, and therefore, as in the classical case, it is scattered into one arm, and remains in it.
Comment: without imposing the unitarity, Eq.(\ref{xHx}) would also contain terms
 $\sim exp[\pm i(x_2-x_3)^2/(2mt)]$ that would yield generally nonzero contribution also after averaging.

To conclude, using the representation of the current noise as a sum over
single-particle transitions we have shown that the current correlations in
time and their spectra are different \cite{spatial} in a transport and a
beam experiment, although the average current is the same.
Thus, the picture of current in a degenerate Fermi gas as a beam of particles
with energies in the transport window is grossly over-simplified.
 For a three-terminal device, which is a solid state analog of a beam splitting setup
(from arm 1 to arms 2 and 3), the difference is given by the second term
in Eq.(\ref{Sbeam-final}), which exists only in the beam, and which start to be important
at $\omega$ of order of $eV$. In the range $\omega>eV$,
 where there is no noise in the transport gas, this
extra term gives Poissonian white noise for the auto-correlators
$S_{22}(\omega)$ and $S_{33}(\omega)$, but does \textit{not} contribute to the
cross-correlator $S_{23}(\omega)$.

This project was supported by the Israel Science Foundation, by the
German-Israeli Foundation and by a joint grant from the Israeli
Ministry of Science and the French Ministry of Research and Technology.
We thank E. Comforti, B. Dou\c{c}ot, C. Glattli, M. Heiblum and D. Prober for instructive
discussions. Some of this work was done while YI and UG were visiting the
Ecole Normale Superieure in Paris. They acknowledge support by the Chaires
Internationale Blaise Pascal.

\end{document}